\documentclass[12pt,a4paper,showkeys]{revtex4}
\usepackage{graphicx}
\usepackage{epsfig}
\usepackage{bm}
\usepackage{amsmath}
\usepackage{amssymb}
\usepackage{graphics}
\usepackage{epstopdf}
\usepackage[linkcolor=red,citecolor=green,urlcolor=blue]{hyperref}

\begin{document}

\title{ \textbf{An update on the evolution of double parton distributions}}
\author{Federico Alberto Ceccopieri}
\email{federico.alberto.ceccopieri@cern.ch}
\affiliation{IFPA, Universit\'e de Li\`ege,  All\'ee du 6 ao\^ut, B\^at B5a,   \\4000
Li\`ege, Belgium}
\begin{abstract}
We consider double parton distributions in the general case in which 
the virtualities of the interacting partons are different. We elaborate 
the corresponding evolution equations and their extension to next-to-leading logarithmic accuracy.
\end{abstract}


\maketitle

\section{Introduction}
\noindent
In hadron-hadron collisions it is often assumed that final states containing 
high-mass systems or high transverse momentum jets are generated by 
a single hard scattering which involves one parton 
from each colliding hadron. The possibility, however, of multiple hard scatterings 
should be considered as well. One might therefore consider 
the case in which two hard interactions occur within the same hadron-hadron collision 
as an approximation to the full multiple parton interactions contributions.
Several experimental results indeed support this possibility
and are based on the analysis of the four-jet~\cite{AFS,UA2,cdf_4jets}
and $\gamma$ + 3 jets channels~\cite{cdf,D0}.
Multiple parton interactions has been first modelled and included in modern 
Monte Carlo event generators~\cite{MPI,sjostrand2,sjostrand3}. 
Very recently detailed phenomenological investigations on double parton scattering 
have appeared in the literature. 
They focus on the four-jet~\cite{BDFD,BJS_4jet}, 
double-inclusive-forward-pion production~\cite{double_forward_pion}, 
same-sign $W$~\cite{same-signW} and $Z$ plus jets~\cite{maina} final states.
The efforts to identify processes which could be maximally sensitive to the contributions
of double-parton scattering (DPS) is driven by two main interests. 
On the one hand a carefull assessment of phase-space region where 
DPS events might impact searches for new physics is needed. 
On the other hand a genuine understanding of hadron structure 
in high energy collisions in terms of multi-partons distributions 
would emerge from these studies. 
Most of the predictions reported in the phenomenological analysis are based on the simplified model in which double parton distributions (DPD) are supposed to be the product of single-parton distributions. This assumption is indeed reasonable 
given the regime of low parton fractional momenta presently accessible
at hadron colliders. Such an assumption simply disregards any longitudinal-momentum 
and flavour correlation between the two interacting partons from each hadron, so that each one evolves according 
to standard DGLAP equations~\cite{DGLAP}. The main virtue of such an approach is that it is
technically appealing since numerous single parton distributions sets are available. 
 The scale dependence of double-parton distributions 
has been worked out in Ref.~\cite{snigirev2003}. With respect to standard single-parton distributions evolution equations (DGLAP), they do contain an additional 
term which is responsable for dynamical correlation between the interacting partons.
Quite recently a new set of double parton distributions has been obtained by means of numerical 
integrations of the DPD evolutions equations. The initial conditions are such that DPD 
preserve under evolution a number of momentum and flavour sum rules~\cite{GS09}. 
The evolution equations elaborated in Ref.~\cite{snigirev2003} however assume that 
both the interacting parton have the same virtualities.
   
Numerical studies~\cite{maina,BJS_4jet} and the arguments given in Ref.~\cite{Diehl} indeed indicate that the characterizing scale for double parton scattering 
is the transverse momentum of the final state products.  
One may therefore consider the production of 
a gauge boson of mass $M^2 = Q_2^2$ in the first hard scattering associated with jets
produced in the second hard scattering and   
characterized by the jet transverse momentum $P_t^2 = Q_1^2$. 
We indicate with $Q_1^2$ and $Q_2^2$ the factorization scales for the two hard processes. 
The low $P_t^2$ regime, with $P_t^2 \ll M^2$, for which we expect significant contributions from 
DPS events, is not covered by evolution equations proposed in Ref.~\cite{snigirev2003}.
The first purpose of this paper is to obtain DPD evolutions equations for different virtualities
of the interacting partons.
Then we consider the extension of the formalism beyond the leading logarithmic approximation. 
By using jet calculus rules we work out the inhomogeneous term at next-to-leading 
order accuracy and connect the real two-loops splitting functions
arising in DPD evolution equations to the one appearing in fracture functions 
evolution equations at the same level of accuracy.
Our main results are all framed within the Jet Calculus formalism since it proves to be 
an efficient tool for calculating multi-parton distributions properties and 
only an \textsl{ab initio} calculation could bring these findings on a firmer ground.\\

This paper is organized as follows. In Sec.~\ref{DPD_intro} 
we review the basics of Jet Calculus formalism and recover known results on DPD.
In Sec.~\ref{DPD_at_different_q} we work out 
the DPD evolution equations at different virtualities. 
In Sec.~\ref{DPD_NLL} we guess the evolution equations for DPD at next-to-leading order accuracy. 
Finally we summarise our results in Sec.~\ref{summary}.

\section{Preliminaries}
\label{DPD_intro}
\begin{figure}[h]
\includegraphics[width=13cm,height=5cm,angle=0]{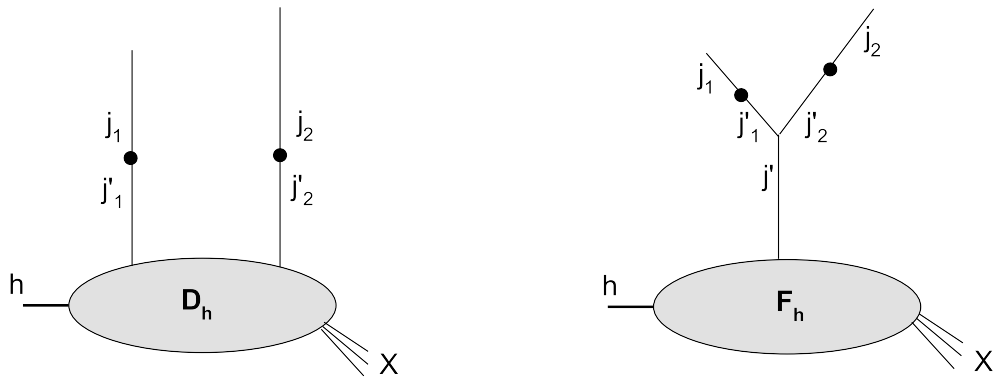}
\caption{Pictorial representation of both terms on right hand side of eq.~(\ref{Ddef}).
Black dots symbolize the parton-to-parton evolution function, $E$\,. }
\label{fig1}
\end{figure}

The double-parton distributions (DPD) $D_h^{j_1,j_2}(x_1,Q_1^2,x_2,Q_2^2)$
are interpreted as the two-particle inclusive probability of finding in a target hadron a couple of partons
of flavour $j_1$ and $j_2$, fractional momenta $x_1$ and $x_2$ and
virtualities up to $Q_1^2$ and $Q_2^2$, respectively. 
The special case in which 
$Q_1^2=Q_2^2=Q^2$ has been considered in detail in Ref.~\cite{snigirev2003}.
According to Jet Calculus~\cite{KUV}, the distributions at the final scales,
$Q_1^2$ and $Q_2^2$, are constructed through the parton-to-parton 
functions, $E$, which themselves obey DGLAP-type~\cite{DGLAP} evolution equations:
\begin{equation}
\label{Eevo}
Q^2 \frac{\partial}{\partial Q^2} E_i^j(x,Q_0^2,Q^2)=\frac{\alpha_s(Q^2)}{2\pi}
\int_x^1 \frac{du}{u} P_k^i(u) E_i^k(x/u,Q_0^2,Q^2)\,,
\end{equation}
where $P_k^i(u)$ are the Altarelli-Parisi splitting functions.
Inserting the initial condition \\ $E_i^j(x,Q_0^2,Q^2)=\delta_i^j \delta(1-x)$ eq.~(\ref{Eevo})
can iteratively be solved to give
\begin{equation}
\label{Eexpansion}
E_i^j(x,Q_0^2,Q^2)=\delta_i^j \delta(1-x)+\frac{\alpha_s}{2\pi} P_i^j(x) \ln\frac{Q^2}{Q_0^2}+\mathcal{O}(\alpha_s^2)\,.
\end{equation}
Therefore the functions $E$ provide the resummation of collinear logarithms 
up to the accuracy with which the  $P_k^i(u)$ are specified.
We may therefore express, by Jet Calculus rules~\cite{KUV}, the double-parton distributions $D_h^{j_1,j_2}(x_1,Q_1^2,x_2,Q_2^2)$ as
\begin{eqnarray}
\label{Ddef}
&& D_h^{j_1,j_2}(x_1,Q_1^2,x_2,Q_2^2)=\\
&& \int_{x_1}^{1-x_2} \frac{dz_1}{z_1} \int_{x_2}^{1-z_1} \frac{dz_2}{z_2} 
D_h^{j_1',j_2'}(z_1,Q_0^2,z_2,Q_0^2) 
E_{j_1'}^{j_1}\Big( \frac{x_1}{z_1},Q_0^2,Q_1^2\Big) 
E_{j_2'}^{j_2}\Big( \frac{x_2}{z_2},Q_0^2,Q_2^2\Big) +\nonumber\\
&&\int^{Min(Q_1^2,Q_2^2)}_{Q_0^2} d\mu_s^2  
\int_{x_1}^{1-x_2} \frac{dz_1}{z_1} \int_{x_2}^{1-z_1} \frac{dz_2}{z_2}
D_{h,corr}^{j_1',j_2'}(z_1,z_2,\mu_s^2) 
E_{j_1'}^{j_1}\Big( \frac{x_1}{z_1},\mu_s^2,Q_1^2\Big) 
E_{j_2'}^{j_2}\Big( \frac{x_2}{z_2},\mu_s^2,Q_2^2\Big)\,. \nonumber
\end{eqnarray}
The first term on r.h.s., usually addressed as the homogeneous term,
takes into account the uncorrelated evolution of the active partons found at a scale $Q_0^2$ in $D_h^{j_1',j_2'}$ 
up to $Q_1^2$ and $Q_2^2$, respectively.
The second term, the inhomogeneous one, takes into account the probability to find the active partons 
at $Q_1^2$ and $Q_2^2$ as a result of a splitting at a scale $\mu_s^2$, integrated over all the intermediate
scale at which such splitting may occur. The distribution $D_{h,corr}^{j_1',j_2'}$ is 
\begin{equation}
\label{Dcorr}
D_{h,corr}^{j_1',j_2'}(z_1,z_2,\mu_s^2)=
\frac{\alpha_s(\mu_s^2)}{2\pi \mu_s^2}
\frac{F_h^{j'}(z_1+z_2,\mu_s^2)}{z_1+z_2} 
\widehat{P}_{j'}^{j_1',j_2'} \Big( \frac{z_1}{z_1+z_2} \Big)\,.
\end{equation}
The distributions $F_h^{j'}$ in eq.~(\ref{Dcorr}) are the single parton distributions 
and the $\widehat{P}_{j'}^{j_1',j_2'}$ are the real
Altarelli-Parisi splitting functions~\cite{KUV}.
Both terms in eq.~(\ref{Ddef}) are shown in Fig.~(\ref{fig1}).

\noindent
Due to strong ordering in parton virtualities, the maximum scale in the $\mu_s^2$ integral 
is set to $Min(Q_1^2,Q_2^2)$.
The scale $Q_0^2$ is in general the (low) scale at which DPD are usually modelled, in complete 
analogy with the single-parton distributions case. In the present context it also acts as the factorization 
scale for the correlated term, since all unresolved splittings for which $\mu_s^2<Q_0^2$ are effectively 
taken into account in the definition of $D_h^{j_1',j_2'}(z_1,Q_0^2,z_2,Q_0^2)$.  
The limits on convolutions integrals in eq.~(\ref{Dcorr}) are fixed by momentum conservation,
\begin{equation}
\label{kine}
z_1\ge x_1, \;\;\;  \;\;\;z_2\ge x_2, \;\;\; \;\;\; z_1+z_2\le 1\,,
\end{equation} 
where $z_1$ and $z_2$ are intermediate partons fractional momenta and the last condition guarantees that
their sum never exceeds the incoming hadron fractional momentum. The, lowest-order, real Altarelli-Parisi splitting functions $\widehat{P}_{q}^{qg}(u)$ and $\widehat{P}_{g}^{gg}(u)$ both contain an infrared singularity 
at the endpoint, $u=1$. It is however easy to show that such a singularity is always outside 
the triangle defined by eq.~(\ref{kine}) in the $[z_1,z_2]$ plane, provided that the trivial condition $x_1,x_2>0$ holds.  
In the ``equal scales'' case, $Q_1^2=Q_2^2=Q^2$, we may take the logarithmic 
derivative with respect to $Q^2$ in eq.~(\ref{Ddef}) and recover the result presented in Ref.\cite{snigirev2003}:
\begin{multline}
\label{snigirev_LL}
Q^2\frac{\partial D_h^{j_1,j_2}(x_1,x_2,Q^2)}{\partial Q^2}=
\frac{\alpha_s(Q^2)}{2\pi}
\int_{\frac{x_1}{1-x_2}}^{1}\frac{du}{u} 
P_{k}^{j_1}(u) D_h^{j_2,k}(x_1/u,x_2,Q^2) + \\
\frac{\alpha_s(Q^2)}{2\pi}
\int_{\frac{x_2}{1-x_1}}^{1}\frac{du}{u} 
P_{k}^{j_2}(u) D_h^{j_1,k}(x_1,x_2/u,Q^2) +
\frac{\alpha_s(Q^2)}{2\pi}
\frac{F_h^{j'}(x_1+x_2,Q^2)}{x_1+x_2} \widehat{P}_{j'}^{j_1,j_2}
\Big( \frac{x_1}{x_1+x_2} \Big)\,.
\end{multline}
The first and second terms on the right-hand side are obtained through 
the $Q^2$ dependence contained in the $E$ functions, while the last 
is obtained from the $Q^2$ dependent limit in the $\mu_s^2$ integration
in the correlated term. The evolution equations therefore resum 
large contributions of the type $\alpha_s \ln (Q^2/Q_0^2)$ and 
$\alpha_s \ln (Q^2/\mu_s^2)$ appearing in the uncorrelated and correlated term 
of eq.~(\ref{Ddef}), respectively.
\section{Evolution equations for different virtualities}
\label{DPD_at_different_q}
Let us now consider the general case in which the partons initiating the 
two separate hard scatterings have different virtualities, $Q_1^2$ and $Q_2^2$, respectively with $Q_1^2<Q_2^2$. 
The evolution equations for the higher scale is obtained by taking the logarithmic derivative of
eq.~(\ref{Ddef}) with respect to $Q_2^2$  
\begin{eqnarray}
&&Q_2^2\frac{\partial D_h^{j_1,j_2}(x_1,Q_1^2,x_2,Q_2^2)}{\partial Q_2^2}=
\Bigg[
\int_{x_1}^{1-x_2} \frac{dz_1}{z_1} \int_{x_2}^{1-z_1} \frac{dz_2}{z_2}
D_h^{j_1',j_2'}(z_1,Q_0^2,z_2,Q_0^2) 
E_{j_1'}^{j_1}\Big( \frac{x_1}{z_1}, Q_0^2,Q_1^2\Big) +\nonumber\\
&&+\int^{Q_1^2}_{Q_0^2} d\mu_s^2 
\frac{\alpha_s(\mu_s^2)}{2\pi \mu_s^2}
\int_{x_1}^{1-x_2} \frac{dz_1}{z_1} \int_{x_2}^{1-z_1} \frac{dz_2}{z_2} 
\frac{F_h^{j'}(z_1+z_2,\mu_s^2)}{z_1+z_2} 
\widehat{P}_{j'}^{j_1',j_2'}
\Big( \frac{z_1}{z_1+z_2} \Big) E_{j_1'}^{j_1}\Big( \frac{x_1}{z_1}, \mu_s^2,Q_1^2\Big)\Bigg]\cdot \nonumber\\
&& \hspace{8cm} \cdot \frac{\alpha_s(Q_2^2)}{2\pi}
\int_{\frac{x_2}{z_2}}^{1}  \frac{du}{u} P_{k}^{j_2} (u) E_{j_2'}^{k}(\frac{x_2}{z_2u},\mu_s^2,Q_2^2)\,,
\end{eqnarray}
and using eq.~(\ref{Eevo}). Reordering the integrals, we get
\begin{eqnarray}
&&Q_2^2\frac{\partial D_h^{j_1,j_2}(x_1,Q_1^2,x_2,Q_2^2)}{\partial Q_2^2}=
\frac{\alpha_s(Q_2^2)}{2\pi}\int_{\frac{x_2}{1-x_1}}^{1}\frac{du}{u} P_k^{j_2}(u) \Bigg[
\nonumber\\
&&\int_{x_1}^{1-\frac{x_2}{u}} \frac{dz_1}{z_1} \int_{\frac{x_2}{u}}^{1-z_1} \frac{dz_2}{z_2}
D_h^{j_1',j_2'}(z_1,Q_0^2,z_2,Q_0^2) 
E_{j_1'}^{j_1}\Big( \frac{x_1}{z_1}, Q_0^2,Q_1^2\Big) 
E_{j_2'}^k \Big(\frac{x_2}{z_2 u},Q_0^2,Q_2^2\Big) +\nonumber\\
&&+\int^{Q_1^2}_{Q_0^2} d\mu_s^2 
\frac{\alpha_s(\mu_s^2)}{2\pi \mu_s^2}
\int_{x_1}^{1-\frac{x_2}{u}} \frac{dz_1}{z_1} \int_{\frac{x_2}{u}}^{1-z_1} \frac{dz_2}{z_2} 
\frac{F_h^{j'}(z_1+z_2,\mu_s^2)}{z_1+z_2} 
\widehat{P}_{j'}^{j_1',j_2'}
\Big( \frac{z_1}{z_1+z_2} \Big) E_{j_1'}^{j_1}\Big( \frac{x_1}{z_1}, \mu_s^2,Q_1^2\Big)\cdot \nonumber\\
&& \hspace{12cm}\cdot E_{j_2'}^k(\frac{x_2}{z_2 u},\mu_s^2,Q_2^2) \Bigg]\,.
\end{eqnarray}
It is now easy to recognize, through direct comparison with eq.(\ref{Ddef}),
that the term is square brackets is the double parton distribution $D_h^{j_1,k}(x_1,Q_1^2,x_2/u,Q_2^2)$.
The desidered evolution equations then becomes
\begin{equation}
\label{evoQ2}
Q_2^2\frac{\partial D_h^{j_1,j_2}(x_1,Q_1^2,x_2,Q_2^2)}{\partial Q_2^2}=
\frac{\alpha_s(Q_2^2)}{2\pi} \int_{\frac{x_2}{1-x_1}}^{1}\frac{du}{u} 
P_{k}^{j_2}(u) D_h^{j_1,k}(x_1,Q_1^2,x_2/u,Q_2^2)\,.
\end{equation}
We could obtain the same result in a rather different way. 
We can in fact exploit the following property of the $E$ function
\begin{equation}
\label{Egroup}
E_i^j(x,Q_0^2,Q_2^2)=\int_x^1 \frac{du}{u} E_i^k\Big( \frac{x}{u},Q_0^2,Q_1^2\Big)
E_k^j(u,Q_1^2,Q_2^2)\,. 
\end{equation}
The latter can be checked, for example, by expanding the $E$ functions in power of $\alpha_s$
as given in  eq.~(\ref{Eexpansion}).
By using eq.~(\ref{Egroup}), eq.~(\ref{Ddef}) can be recast in the much compact form   
\begin{equation}
\label{fast}
D_h^{j_1,j_2}(x_1,Q_1^2,x_2,Q_2^2)=\int_{x_2}^{1-x_1} \frac{dw_2}{w_2} 
D_h^{j_1,k}(x_1,Q_1^2,w_2,Q_1^2) 
E_{k}^{j_2}\Big( \frac{x_2}{w_2}, Q_1^2,Q_2^2\Big) \,.
\end{equation} 
By direct substitution it can be checked that eq.~(\ref{fast}) is indeed a solution of eq.~(\ref{evoQ2}). 
With respect to ``equal scale" DPD evolution equations we notice the disappereance
of the inhomogenous term. This is due to the fact that 
the correlations up to a scale $Q_1^2$ given by the inhomogeneous term are taken into account by the ``equal scales" evolution equations and properly built into $D_h^{j_1,k}(x_1,Q_1^2,w_2,Q_1^2)$ . The evolution of the second parton from $Q_1^2$ to $Q_2^2$
is uncorrelated due to strong ordering in virtualities assumed in the leading logarithmic approximation. 
From the numerical point of view therefore DPD at different virtualities  
can be obtained evolving $D_h^{j_1,k}(x_1,Q_1^2,w_2,Q_1^2)$ with the ``equal scale" evolution 
equations up to $Q_1^2$, eq.~(\ref{snigirev_LL}), and then using the latter output 
as initial condition in eq.~(\ref{evoQ2}), for $Q_2^2>Q_1^2$. 
We have threfore proven the conjecture put forward in Ref.~\cite{GS09} and 
actually implemented numerically \cite{GS09_web}.
For completeness we have also considered 
the DPD evolution equations in $Q_1^2$. Provided that $Q_1^2<Q_2^2$ and using the same techniques
through which we have derived eqs.~(\ref{snigirev_LL}) and (\ref{evoQ2}) we get
\begin{multline}
\label{evoQ1}
Q_1^2\frac{\partial D_h^{j_1,j_2}(x_1,Q_1^2,x_2,Q_2^2)}{\partial Q_1^2}=
\frac{\alpha_s(Q_1^2)}{2\pi} \int_{\frac{x_1}{1-x_2}}^{1}\frac{du}{u} 
P_{k}^{j_1}(u) D_h^{k,j_2}(x_1/u,Q_1^2,x_2,Q_2^2)+\\
+\frac{\alpha_s(Q_1^2)}{2\pi} \int_{x_2}^{1-x_1} \frac{dz_2}{z_2}
\frac{F_h^j(x_1+z_2,Q_1^2)}{x_1+z_2} \widehat{P}_{j}^{j_1 j_2'} 
\Big( \frac{x_1}{x_1+z_2} \Big) E_{j_2'}^{j_2} 
\Big( \frac{x_2}{z_2},Q_1^2,Q_2^2 \big)\,.
\end{multline}
In this case the evolution equations contain an inhmogeneous term which
arises due to the explicit $Q_1^2$ dependence on the $\mu_s^2$ integral in eq.~(\ref{Ddef}). 
Since the factorization scale are kept different, the latter does contain 
explicitely the function $E(Q_1^2,Q_2^2)$, which cannot be further simplified.
To avoid a direct calculations of the $E$ function, the double-parton distributions 
for unequal final scales should be obtained therefore via the two step procedure mentioned above.

\section{Evolution equations to NLLA}
\label{DPD_NLL}
In this section we address the problem of deriving the structure of DPS evolution equations 
at next-to-leading logarithmic accuracy. The aim therefore is to provide some guidance for 
an eventual \textsl{ab initio} calculation. At present, in fact, such an accuracy is not required
since, given the scarce experimental information available, 
we do not even have sufficient data to test whether the scale dependence predicted by DPD evolution is supported.
Jet Calculus techniques has been succesfully extended up next-to-leading
logarithmic accuracy to improve the perturbative description of time-like parton cascades~\cite{jet_NLL}.
For space-like parton cascades instead, which is the case we are actually interested in, the formalism has not been 
extended beyond leading-logarithmic accuracy. However a couple of calculations have been performed
in the context of semi-inclusive Deep Inelastic Scattering. In particular the one-particle inclusive 
cross sections up to order $\mathcal{O}(\alpha_s^2)$ have been calculated in Refs.~\cite{fracture2loop_quark,fracture2loop_gluon}. Such calculations carefully consider 
hadron production collinear to the hadron remnant where the introduction of fracture functions~\cite{Trentadue_Veneziano} is shown 
to be necessary to factorize additional collinear singularities appearing in the calculations in that phase-space region.
The fixed order calculations at $\mathcal{O}(\alpha_s^2)$ allows the authors to derive the fracture functions evolution equations to next-to-leading logarithmic accuracy, as well as the two-loop, unknown, 
real splitting functions, $\widehat{P}^{(1)}$.
Fracture functions evolution equations can be calculated, as DPD, within the 
Jet Calculus formalism~\cite{extendedM,newfracture} and  they do contain an inhomogenous term as well.  
While in the fracture functions case the partons emitted by the active one hadronizes 
through a fragmentation function, in the DPD one, the emitted parton is allowed to further evolve and eventually initiate a second hard scattering.   

When evaluting the evolutions equations 
at next-to-leading logarithmic accuracy the evolution equations for the 
parton-to-parton functions $E$ must be properly modified to 
\begin{equation}
\label{EevoNLL}
Q^2 \frac{\partial}{\partial Q^2} E_i^j(x,Q_0^2,Q^2)=\frac{\alpha_s(Q^2)}{2\pi}
\int_x^1 \frac{du}{u} \Big[P_k^{(0),i}(u)+ \frac{\alpha_s(Q^2)}{2\pi} P_k^{(1),i}(u) 
\Big]E_i^k(x/u,Q_0^2,Q^2)\,,
\end{equation}
where $P^{(0)}(u)$ and $P^{(1)}(u)$ are the one- and two-loops~\cite{AP2loop} Altarelli-Parisi splitting functions,
respectively. This in turn implies that the two homogenous terms in DPD evolution
equations in eq.~(\ref{snigirev_LL}) are modified by adding the two-loop splitting functions contributions. 
On the contrary, the derivation of the inhomogenous term to next-to-leading logarithmic accuracy is not trivial 
so, in the following, we will construct it explicitely in the ``equal scales" case. The 
correlated term can be written therefore as 
\begin{multline}
\label{Dcorr3}
D_{h,corr}^{j_1',j_2'}(x_1,x_2,Q_0^2,Q^2)=
\int_{Q_0^2}^{Q^2} \frac{\alpha_s(\mu_s^2)}{2\pi \mu_s^2}
\int_{x_1+x_2}^{1} dw  
\int_{x_1}^{1-x_2} \frac{dz_1}{z_1} \int_{x_2}^{1-z_1} \frac{dz_2}{z_2}
\int dr_1 \;dr_2 \;du_1 \;du_2 \\
 \cdot F_h^{j'}(w,\mu_s^2)) \Big[ \widehat{P}_{j'}^{(0)\,j_1',j_2'}(u_1) \delta(1-u_1-u_2)
+ \frac{\alpha_s(\mu_s^2)}{2\pi \mu_s^2}
\widehat{P}_{j'}^{(1)\,j_1',j_2'}(u_1,u_2)\Big]  \cdot\\
  \cdot E_{j_1'}^{j_{1}}(r_1,\mu_s^2,Q^2) \; E_{j_2'}^{j_2}(r_2,\mu_s^2,Q^2) 
\;\delta(x_1-r_1 z_1)\;\delta(x_2-r_2 z_2)\;\delta(z_1-u_1 w)\;\delta(z_2-u_2 w)\,.
\end{multline}
In the above equations $\widehat{P}_{j'}^{(1)\,j_1',j_2'}(u_1,u_2)$ gives the probability 
that a parton $j'$ splits to three partons, where the first, $j_1'$, and a second, $j_2'$, have respectively 
a fraction $u_1$ and $u_2$ of the incoming parton momentum $j'$ and the third is integrated over.  
Integrating the $\delta$-functions, which implements longitudinal momentum conservation, one gets
\begin{multline}
\label{Dcorr2}
D_{h,corr}^{j_1',j_2'}(x_1,x_2,Q_0^2,Q^2)=
\int_{Q_0^2}^{Q^2} \frac{\alpha_s(\mu_s^2)}{2\pi \mu_s^2}
\int_{x_1+x_2}^{1} \frac{dw}{w^2} F_h^{j'}(w,\mu_s^2)
\int_{x_1}^{1-x_2} \frac{dz_1}{z_1} \int_{x_2}^{1-z_1} \frac{dz_2}{z_2} \\
 \Big[ \widehat{P}_{j'}^{(0)\,j_1',j_2'}\Big(\frac{z_1}{w}\Big) 
\delta\Big(1-\frac{z_1}{w}-\frac{z_2}{w} \Big)
+ \frac{\alpha_s(\mu_s^2)}{2\pi \mu_s^2}
\widehat{P}_{j'}^{(1)\,j_1',j_2'}\Big(\frac{z_1}{w},\frac{z_2}{w}\Big)\Big]\\
E_{j_1'}^{j_{1}}\Big(\frac{x_1}{z_1},\mu_s^2,Q^2\Big) \; E_{j_2'}^{j_2}
\Big( \frac{x_2}{z_2},\mu_s^2,Q^2\Big) \,.
\end{multline}
As already noted, the inhomogenous term in DPD evolution equations is due  
to the explicit $Q^2$ dependence in the upper limit of $\mu_s^2$ integration.
In order to obtain it we  set $\mu_s^2=Q^2$ in eq.(\ref{Dcorr2}), multiply by $Q^2$, and use intial condition on $E$, 
$E_i^j(x,Q^2,Q^2)=\delta_i^j \delta(1-x)$. Adding the homogeneous contributions, the final result reads
\begin{multline}
\label{DPS_NLL}
Q^2\frac{\partial D_h^{j_1,j_2}(x_1,x_2,Q^2)}{\partial Q^2}=
\frac{\alpha_s(Q^2)}{2\pi}
\int_{\frac{x_1}{1-x_2}}^{1}\frac{du}{u} 
\Big[ P_{k}^{(0),j_1}(u) + \frac{\alpha_s(Q^2)}{2\pi} P_{k}^{(1),j_1}(u) \Big]
 D_h^{k,j_2}(x_1/u,x_2,Q^2) + \\
+\frac{\alpha_s(Q^2)}{2\pi}
\int_{\frac{x_2}{1-x_1}}^{1}\frac{du}{u} 
\Big[ P_{k}^{(0),j_2}(u) + \frac{\alpha_s(Q^2)}{2\pi} P_{k}^{(1),j_2}(u) \Big]
D_h^{j_1,k}(x_1,x_2/u,Q^2) +\\
+\frac{\alpha_s(Q^2)}{2\pi} 
\int_{x_1+x_2}^1 \frac{dw}{w^2}F_h^{j'}(w,Q^2)
\Big[
w \widehat{P}_{j'}^{(0),j_1,j_2} \Big( \frac{x_1}{w} \Big) \delta(w-x_1-x_2) 
+\frac{\alpha_s(Q^2)}{2\pi} \widehat{P}^{(1),j_1,j_2}_{j'} \Big( \frac{x_1}{w},\frac{x_2}{w}\Big)
\Big]\,.
\end{multline}
It should be noted however that the kernels $\widehat{P}'^{(1),j_1',j_2'}_{j'}(u,v)$
reported in Refs.~\cite{fracture2loop_quark,fracture2loop_gluon}
do express the probability that a parton $j'$ splits into a parton $j_1'$ with a momentum fraction 
$u$ of the incoming parton,  into a parton $j_2'$ with a mometum fraction $v$ of $j_1'$,
the third being integrated over. 
Therefore they are related to the ones appearing in eq.~(\ref{DPS_NLL})
by the following mapping 
\begin{equation}
\widehat{P}_{j'}^{(1),j_1',j_2'}(u_1,u_2) =\frac{1}{u_1}  \widehat{P}'^{(1),j_1',j_2'}_{j'}
\Big(u_1,\frac{u_2}{u_1}\Big)\,.
\end{equation}
The additional integral in the inhomogeneous term does appear since 
the momentum is not anymore constrained in the $1\rightarrow 2$ splitting.
The DPS evolution equations to next-to-leading logarithmic accuracy for different scales 
can be obtained by the same arguments given in Sec.~\ref{DPD_at_different_q}. 
We just quote the final result which reads 
\begin{multline}
\label{evoQ2_NLL}
Q_2^2\frac{\partial D_h^{j_1,j_2}(x_1,Q_1^2,x_2,Q_2^2)}{\partial Q_2^2}=
\frac{\alpha_s(Q_2^2)}{2\pi} \int_{\frac{x_2}{1-x_1}}^{1}\frac{du}{u} 
\Big[ P_{k}^{(0),j_2}(u) + \frac{\alpha_s(Q_2^2)}{2\pi} P_{k}^{(1),j_2}(u) \Big] \cdot \\
\cdot  D_h^{j_1,k}(x_1,Q_1^2,x_2/u,Q_2^2)\,,
\end{multline}
provided that $Q_1^2<Q_2^2$.

\section{Summary}
\label{summary}
\noindent
We have considered double parton distributions in the general case in which 
the two factorization scales are kept different and derivered the corresponding evolution equations.   
The results of the present calculation support 
the guess put forward in Ref.~\cite{GS09} and recently implemented numerically~\cite{GS09_web}
widening the range of possible phenomenlogical investigations on double-parton scatterings.
We have also derived the general structure of the DPD evolution equations
at next-to-leading logarithmic accuracy and indicated how to transform the two-loops real 
splitting functions present in the literature in order to be used  in the present context. 
Both results should be confirmed by performing an \textsl{ab initio} calculation.

\begin{acknowledgments}
The author warmly thanks Jean-Ren\'e Cudell for stimulating discussions on the subject. 
\end{acknowledgments}

\end{document}